\pdfoutput=1

\documentclass[conference]{IEEEtran}
\IEEEoverridecommandlockouts
\usepackage{graphicx}
\usepackage{bbm}
\usepackage{amsfonts}
\usepackage{chngcntr}
\usepackage{times}
\usepackage{latexsym}
\usepackage{amssymb}
\usepackage{amsmath}
\usepackage{cite}
\usepackage{verbatim}
\usepackage{subfigure}
\usepackage{multirow}
\usepackage{lastpage}
\usepackage{calc}
\usepackage{eso-pic}
\usepackage{tabularx}
\usepackage{ifthen}
\usepackage{epstopdf}
\usepackage[shortlabels]{enumitem}
\usepackage{textcomp}
\usepackage{caption}
\usepackage[font={small}]{caption}

\usepackage[english]{babel}
\usepackage{textgreek}
\usepackage[utf8x]{inputenc}
\usepackage{algorithm}
\usepackage{algorithmic}

\newtheorem{lemma}{Lemma}

\def\bb0{{\mathbb{0}}}

\def\bb{{\mathbf{b}}}

\def\b0{{\mathbf{0}}}

\def\bbE{{\mathbb{E}}}

\def\bbP{{\mathbb{P}}}

\def\bbR{{\mathbb{R}}}

\def\sf0{{\mathsf{0}}}

\usepackage{footnote}
\allowdisplaybreaks

\ifCLASSINFOpdf
\else
\fi
\IEEEoverridecommandlockouts
\IEEEpubid{\makebox[\columnwidth]{ 978-1-5386-8380-4/19/\$31.00~
		\copyright2019
		IEEE \hfill} \hspace{\columnsep}\makebox[\columnwidth]{ }}
\begin{document}

\title{Device-to-Device Communications in Millimeter Wave Band:  Impact of Beam Alignment Error}

\author{\IEEEauthorblockN{Niloofar Bahadori\IEEEauthorrefmark{1}, Nima Namvar\IEEEauthorrefmark{1}, Brian Kelley\IEEEauthorrefmark{2}, Abdollah Homaifar\IEEEauthorrefmark{1}
	}
%\IEEEauthorblockA{Department of Electrical and Computer Engineering }
%Eldon Tyrell\IEEEauthorrefmark{4} \IEEEauthorrefmark{1}
\IEEEauthorblockA{ North Carolina A\&T State University }
\IEEEauthorblockA{\IEEEauthorrefmark{2} University of Texas at San Antonio
\\
Email:\{nbahador, nnamvar\}@aggies.ncat.edu}brian.kelley@utsa.edu, homaifar@ncat.edu
}

\maketitle

\begin{abstract}
Exploiting the millimeter wave (mmWave) band has recently attracted considerable attention as a potential solution to widespread deployment of device-to-device (D2D) communication challenges, namely, spectrum scarcity and interference. However, its directional nature makes the utilization of mmWave band a challenging task as it requires careful beam alignment between the D2D transmitter and receiver. In this paper, we investigate the impact of inaccurate angle-of-arrival (AoA) estimation as a beam alignment impairment on the performance of a directional mmWave D2D network. We have used tools from stochastic geometry to quantify the signal-to-interference-plus-noise ratio (SINR) coverage probability in the presence of beam misalignment, which can be applied to evaluate D2D network performance. Moreover, the analytical results are verified to be reliable and effective through extensive simulations. Finally, the coverage probability of the D2D network with erroneous beam alignment is compared to the network with perfect beam alignment. The numerical results indicate that the beam misalignment can lead to significant losses in the network's coverage probability.

\emph{Keywords}- Millimeter wave; Device-to-Device communication; Beam misalignment
\end{abstract}
%****************************************************************************
%****************************************************************************
%************************  Introduction  ************************************
%****************************************************************************
%****************************************************************************
\section{Introduction}

Device-to-device (D2D) communication is envisioned to be an integral part of the fifth generation (5G) cellular networks \cite{asadi2014survey}.
Despite its potential advantages, the large-scale implementation of D2D communication has been delayed mainly due to spectrum scarcity in sub-6 GHz band, which leads to severe multi-user interference (MUI) \cite{tehrani2014device}. Utilizing the abundant unlicensed bandwidth in the millimeter wave (mmWave) band is seen as a promising candidate for addressing D2D communications impediments \cite{qiao2015enabling}.

%Device-to-device (D2D) communication, as an integral part of the fifth generation (5G) cellular network, is envisioned to increase the network spectral efficiency and reduce the power consumption \cite{asadi2014survey}.
%In spite of its potential advantages, the large-scale implementation of D2D communications has not been realized, mainly due to spectrum scarcity in the microwave band, which leads to severe multi-user interference (MUI) \cite{tehrani2014device}. Utilizing the abundant unlicensed bandwidth in the millimeter wave (mmW) band is seen as a promising candidate for addressing D2D communications challenges \cite{qiao2015enabling}.

Radio propagation at mmWave band encounters several obstacles such as sever path-loss and sensitivity to blockage \cite{rappaport2013millimeter}. The small wavelength of mmWave signals, however, facilitates implementation of large directional and high-gain antenna arrays on D2D devices, which helps to compensate for additional path-loss \cite{wei2014key}. This, in turn, introduces a new challenge to D2D communication, as achieving the maximum directivity gain in a highly directional mmWave band system requires the transmitter and receiver to be precisely aligned.

In the implementation of directional antennas, the angle-of-arrival (AoA), which represents the angle of incidence of incoming signal power, is acquired to enable D2D devices to steer and align their antenna bore-sight toward the desired signal. In practice, the AoA estimation is not completely accurate due to multiple sources of error such as antenna configuration perturbations and mobility of devices \cite{li2006outage}. Any error in estimating the AoA leads to beam alignment error, which subsequently may cause significant array gain variation or even signal outage at the receiver. Hence, it is crucial to analyze the impact of beam alignment error on the performance of the mmWave D2D network.

%Therefore, in practice the maximum array gain cannot be achieved due to the presence of multiple sources of inaccuracies in AoA estimation such as antenna configuration perturbations and mobility of devices \cite{li2006outage}. Any error in estimating the AoA may lead to a large array gain variation or even complete signal outage at the receiver. Therefore, it is necessary to analyze the impact of beam alignment error on the performance of the network.

Nevertheless, a majority of research work in the area of directional mmWave band communication assumes perfect beam alignment \cite{bai2015coverage,thornburg2016performance,bahadori2018device}. We note several works that explicitly account for alignment error in directional wireless networks \cite{li2006outage,yang2016analysis,wildman2014joint,thornburg2015ergodic}. Authors in \cite{li2006outage}, investigated the impact of an erroneous uniform linear array beamformer on network outage probability and reported a degradation in the network's performance in the presence of beamforming error. A stochastic geometry framework is used in \cite{wildman2014joint} to capture the effect of beam misdirection using the flat-top antenna model. The loss in the capacity and signal power due to misalignment in a mmWave band directional communication network is quantified in \cite{thornburg2015ergodic} and \cite{yang2016analysis}, respectively. However, most of the existing analyses are either performed in sub-6 GHz band or failed to consider the sensitivity of mmWave band communication to blockage. For analytical tractability, some works adopted the simplified flat-top antenna model, which fails to provide an accurate model for network assessment. Others merely assumed small alignment errors for tractability, which is not a valid assumption for mmWave band D2D network due to the users' mobility. We are aware of no work that considers the impact of erroneous beam alignment while taking into account all of the mentioned gaps simultaneously.

In this paper, we analyze the impact of inaccurate AoA estimation on the coverage probability of a directional mmWave D2D network, where the network elements are modeled as a homogeneous Poisson point process (PPP).
The directional antenna is approximated by adopting the cosine antenna model, which compared to the simplified flat-top model, provides a better approximation for the antenna pattern.
The erroneous beam alignment due to AoA estimation is characterized by uniform and Gaussian distributions. We have used tools from stochastic geometry to derive the distribution of the received signal and interference, which is used to quantify the network performance. The analytical results are shown to be computationally precise through numerical evaluation. Lastly, in order to assess the impact of alignment error, the performance of the D2D network with beam misalignment is compared to the one with perfect beam alignment. Simulation results show that network performance in term of coverage probability can be affected significantly by beam misalignment.

\setlength\belowcaptionskip{-2.45ex}
\begin{figure}
	\centering
	\includegraphics[width=7.5cm,height=4.5cm, trim=1.5cm 8.6cm 1.5cm 9.0cm, clip]{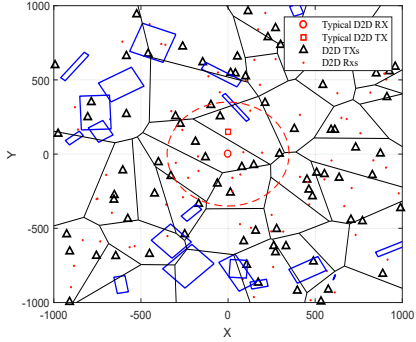}
	\caption{A sample realization of the described network model, where blue rectangles represent the building blockages.}
	\label{fig:network}
\end{figure}
The remainder of this paper is organized as follows. The system model and performance measure are described in Section \ref{sec:system model}. The beam misalignment along with the directional antenna gain distribution in the presence beam alignment error is characterized in Section \ref{sec:antModel}. Then the coverage probability, as well as the impact of antenna misalignment for the mmWave D2D network, are derived in Section \ref{sec:coverageProb} using stochastic geometry. Numerical results are presented in Section \ref{sec:result} and finally, conclusions are drawn in Section \ref{sec:Conclusion}.
%%%%%%%%%%%%%%%%%%%%%%%%%%%%%%%%%%%%%%%%%%%%%%%%%%%%%%%%%%%%%%%%%%%%%%%%%%%%%%%%%%%%%%%%%%%%%%%%%%%%%%%%%%%%%%%%%%%%%%%%%%%%%%%%%%%%%%%%%%%%%%%%%%%%%%%%%%%%
%%%%%%%%%%%%%%%%%%%%%%%%%%%%%%%%%%%%%%%%%%%%%%%%%%%%%%%%%%%%%%%%%%%%%%%%%%%%%%%%%%%%%%%%%%%%%%%%%%%%%%%%%%%%%%%%%%%%%%%%%%%%%%%%%%%%%%%%%%%%%%%%%%%%%%%%%%%%
%%%%%%%%%%%%%%%%%%%%%%%%%%%%%%%%%%%%%%%%%%%%%%%%%%%%%%%%%%%%%%%%%%%%%%%%%%%%%%%%%%%%%%%%%%%%%%%%%%%%%%%%%%%%%%%%%%%%%%%%%%%%%%%%%%%%%%%%%%%%%%%%%%%%%%%%%%%% width=1\columnwidth

%****************************************************************************
%****************************************************************************
%************************  Methodology  ************************************
%****************************************************************************
%****************************************************************************

%, $\Phi=\{x_i\}$ with density $\lambda$, where $x_i$ denotes the location of $i$-th D2D transmitter

%****************************************************************************
%****************************************************************************
%************************  Analysis  ************************************
%****************************************************************************
%****************************************************************************
\section{System model}\label{sec:system model}
\begin{figure}
	\centering
	\includegraphics[width=.6\columnwidth]{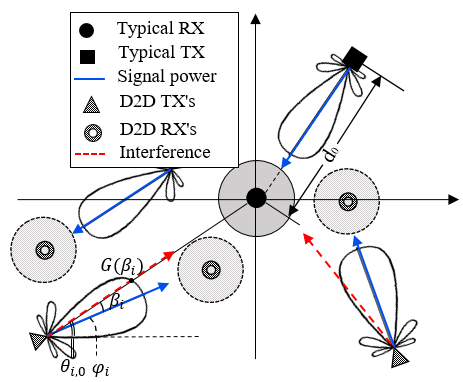}
	\caption{A sample realization of the D2D network with directional transmitter and omni-directional receiver.}
	\setlength{\abovecaptionskip}{-3cm}
	\label{fig:systemModel}
\end{figure}

We consider a directional D2D network in the mmWave band in which D2D transmitters are spatially distributed according to a homogeneous Poisson point process (PPP), denoted as $\mathbf{\Phi}=\{x_i\}$ with density $\lambda$, where $x_i \in \mathbb{R}^2$ denotes the location of $i$-th D2D transmitter. Random size rectangular building blockages are also distributed randomly by another independent PPP. Figure \ref{fig:network} shows a sample realization of the network.
Using the proposed mechanism in \cite{bahadori2018device}, all D2D transmitters are assumed to have a LOS corresponding receiver in its coverage area, and at least one packet ready for transmission.
Without loss of generality, we consider that each D2D receiver has a single receive antenna and its corresponding D2D transmitter is equipped with an array of antennas which allows directional transmission, as depicted in Figure \ref{fig:systemModel}. Since we have no prior information about the D2D transmitter's antenna bore-sight angle, denoted by $\varphi_i$, it is modeled as a uniform random variable as $\varphi\sim \mathcal{U}(-\pi,\pi)$.
Sidestepping the problem of power control, all D2D transmitters are transmitting at a constant transmit power $P_D$. Each communication link experiences i.i.d small-scale Rayleigh fading. Hence, the received signal power can be modeled as an exponential random variable with parameter 1. Moreover, no prior coordination among devices for interference mitigation is assumed.

Here, we use the signal-to-interference-plus-noise ratio (SINR) coverage probability as a metric to assess the performance of the network. The coverage probability is defined as the probability that the received SINR is higher than a predefined threshold $\gamma$, i.e., $p_{\text{c}} (\gamma)=\bbP[\text{SINR}\geq\gamma]$.
The performance metric is obtained for a \emph{typical} D2D transmitter-receiver pair with the receiver located at the origin $(0,0)\in \mathbb{R}^2$, while the result holds for any generic D2D pair, based on the Slivnyak's theorem \cite{baccelli2010stochastic}.

The SINR for the typical receiver can be written as
\begin{equation}
\text{SINR}=\frac{P_D h_0 G_0(\theta) C d_0^{-\alpha}}{\sigma^2+I}\label{eq:SINR},
\end{equation}
where $h_0$ represents the corresponding transmitter's channel gain. Directional antenna gain is parameterized by $G_0(\theta)$, where $\theta$ represents the antenna angle, $C$ symbolizes the path-loss intercept and ${\alpha}$ is the path-loss exponent. The distance between the typical receiver and its transmitter at $x_0$ is denoted by $d_0=\|x_0\|$, and $\|.\|$ represents the Euclidean distance. Finally, $\sigma^2$ and $I$ represent the noise power and aggregate interference, respectively.

\textbf{Blockage model-} In order to model the blockage effect in the mmWave band, we use the line-of-sight (LOS) ball model \cite{bai2015coverage}, in which the actual shape of the LOS region around each D2D receiver is approximated as a fixed-sized ball with radius $R$.
%Hence, all nearby D2D transmitters inside the ball are considered as LOS, while D2D transmitters beyond $R$ are assumed to be non-LOS (NLOS).
Since NLOS transmission in mmWave band suffers from high attenuation, only the interference from LOS D2D transmitters are considered and NLOS transmissions are neglected \cite{rappaport2013millimeter}.
Based on the LOS ball model definition, all LOS D2D transmitters nearby the typical D2D receiver are located inside a disk of radius $R$ centered at the origin.
We note that using the thinning theorem of PPP \cite{baccelli2010stochastic}, LOS D2D transmitters form a PPP, denoted by $\mathbf{\Phi}_L$, with density $\lambda_L$.

The aggregate interference received by the typical D2D receiver can be defined as
\begin{equation}
I = \sum_{i \in \mathbf{\Phi}_L} P_D h_i G_i(\beta_i) C \|x_i\|^{-\alpha}\label{eq:Inter},
\end{equation}
where $h_i$ denotes channel gain of $i$-th LOS D2D transmitter located at $x_i \in \mathbf{\Phi}_L$. The antenna gain of D2D transmitter is characterized by $G_i(\beta_i) $, in which $\beta_i=|\varphi_i-\theta_{i,0}|$, and $\theta_{i,0}$ is the angle between $i$-th transmitter and the typical D2D receiver, as shown in Figure \ref{fig:systemModel}. Note that the antenna angle of interferers, denoted by $\beta_i$, is independent of the location of points in PPP, $\mathbf{\Phi}_L$, and can be considered uniformly distributed as $\varphi_i$ has uniform distribution \cite{wildman2014joint}.

\section{Directional Antenna and Alignment Error}\label{sec:antModel}

\begin{figure}
	\centering
	\includegraphics[width=.57\columnwidth]{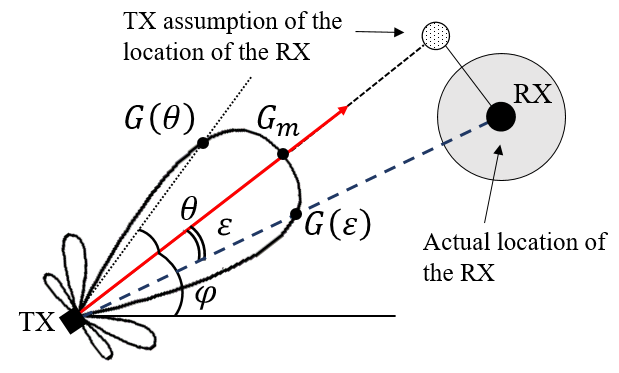}
	\caption{A directional D2D transmitter and its corresponding pair. The red arrow depicts the direction of antennas bore-sight angle, $\varphi$. The D2D transmitter determines the receiver's direction with error of $\varepsilon$ an the antenna gain in the presence of error is $G(\varepsilon)$. }
	\setlength{\abovecaptionskip}{-3cm}
	\label{fig:Angles}
\end{figure}

The directional antenna pattern is modeled using the cosine function. The cosine model provides a better approximation for antenna main-lobe \cite{yu2017coverage}, compared to the flat-top model which is widely used in the literature \cite{bai2015coverage,bai2014analysis,wildman2014joint}. The antenna gain can be defined as
\begin{align}\allowdisplaybreaks
G(\theta)=&
\begin{cases}
G_m\cos^2(\frac{\tau \theta}{2})&  \hspace{2mm} |\theta|\leq \frac{\pi}{\tau} \\
0 & \hspace{2mm} \text{otherwise},
\end{cases}\label{eq:pattern}
\end{align}
where $G_m$ represents the maximum gain, and $\tau$ controls the spread of antenna beam. $\theta$ symbolizes the antenna angle relative to the antenna's bore-sight angle, denoted by $\varphi$, as illustrated in Figure \ref{fig:Angles}.

We assume that each D2D user is enabled to find the direction of its intended peer, using AoA spectrum \cite{bahadori2018device}, and steer its antenna bore-sight toward the direction of its receiver with a simple rotation around its location to transmit the maximum gain. Under the assumption of perfect alignment, each transmitter determines the direction of its receiver accurately. However, accurate beam alignment is not a practical assumption. Any error in estimating the AoA will cause the antenna array to point away from the desired signal and will lead to a reduction of the received power of the desired signal.

\textbf{Beam alignment error-}
Each D2D transmitter determines the AoA (orientation) of its receiver with an additive error, denoted by $\varepsilon_i$.  The alignment error is measured relative to the transmitter's antenna bore-sight angle and characterizes the angle between the actual and estimated bearing of the receiver, as shown in Figure \ref{fig:Angles}.

In this work, the AoA estimation error is characterized by two different random variables, namely, uniform and normal distributions. The uniform distribution is used to model the scenario where our knowledge about the estimation error is limited to the error bounds, and we have no prior information on error magnitude. Therefore, we assume that all error values between the minimum and maximum occur with equal likelihood. On the other hand, the normal distribution is used to model the scenario where we know some values of error are more probable (the ones near the mean value) than others. Note that according to the central limit theorem, the normalized sum of mutually independent random variables with finite variance is well-approximated by normal distribution. Hence, the beam alignment error which stems from multiple independent sources of uncertainty in the system can be modeled as a normal distribution.

\begin{lemma}\label{lem:lemma1}
Given that the AoA estimation error is distributed uniformly, as $\varepsilon \sim \mathcal{U}(-\varepsilon_0, \varepsilon_0)$ with zero mean and $| \varepsilon_0 |< \pi$, the probability distribution function (pdf) of D2D transmitter's antenna gain is
\begin{align}\allowdisplaybreaks
&f_{G}(g)=\frac{1}{\tau \varepsilon_0 \sqrt{g}\sqrt{G_m-g}}\label{eq:pdf},
\end{align}
where
\begin{align}\allowdisplaybreaks
&\text{for} \hspace{2mm}  |\varepsilon_0|<\frac{\pi}{\tau}, \hspace{4mm}   g \in \left[\kappa G_m ,G_m\right]\nonumber\\
&\text{for} \hspace{2mm} \frac{\pi}{\tau}\leq |\varepsilon_0|<\pi , \hspace{4mm}   g \in \left[0 ,G_m\right]\nonumber
\end{align}
and $\kappa=\cos^2(\frac{\tau \varepsilon_0}{2})$.
\end{lemma}
\vspace{2mm}
\begin{IEEEproof}
	 As illustrated in Figure \ref{fig:Angles}, the alignment error reduces the antenna gain to $G(\varepsilon)$, which is less than the antenna maximum gain, $G(\theta=0)=G_m$. In order to characterize the antenna gain distribution, the cumulative distribution function (CDF) of antenna gain can be derived as
	\begin{align}\allowdisplaybreaks
	%\text{for} \hspace{2mm}  |\varepsilon_0|<\frac{\pi}{\tau} \nonumber\\
	F_{G}(g)&=\bbP\left[G(\varepsilon)\leq g\right] \nonumber\\
	&=\bbP\left[ -\varepsilon_0\leq \varepsilon \leq -G^{-1}(g) \right]+\bbP\left[ G^{-1}(g) \leq \varepsilon \leq \varepsilon_0\right]\nonumber\\
	%&\mathrel{\phantom{=}}+\bbP\left[ G^{-1}(g) \leq \varepsilon \leq \varepsilon_0\right]\nonumber\\
	&= \int^{-G^{-1}(g)}_{-\varepsilon_0} \frac{1}{2\varepsilon_0}\text{d}\varepsilon + \int_{G^{-1}(g)}^{\varepsilon_0} \frac{1}{2\varepsilon_0}\text{d}\varepsilon\nonumber\\
	%&\overset{(a)}
	&= 1-\frac{2}{\varepsilon_0\tau}\arccos \left(\sqrt{\frac{g}{G_m}}\right)\nonumber\\
	F_G(g)&=
	\begin{cases}
	0 & g < 0\\
	1-\frac{\pi}{\tau \varepsilon_0} & g = 0 \\
	1-\frac{2}{\varepsilon_0\tau}\arccos \left(\sqrt{\frac{g}{G_m}}\right) & 0 < g \leq G_m\\
	1 & g > G_m \label{eq:CDFCosine}
	\end{cases}
	%\\
	%\text{for} \hspace{2mm} \frac{\pi}{\tau}< |\varepsilon_0|<\pi &\nonumber\\
	%	F_G(g)&=
	%\begin{cases}
	%0 & g < 0\\
	%2(1-\frac{\pi}{ \epsilon_0 \tau}) & g = 0\\
	%1-\frac{2}{\varepsilon_0\tau}\arccos \left(\sqrt{\frac{g}{G_m}}\right) & 0 < g \leq G_m\\
	%1 & g > G_m\label{eq:CDFCosine2}
	%\end{cases}
	\end{align}
	where $G^{-1}(g)=\frac{2}{\tau}\arccos (\sqrt{\frac{g}{G_m}})$. The pdf function in (\ref{eq:pdf}) is derived by taking the derivative of CDF function in (\ref{eq:CDFCosine}).
\end{IEEEproof}
%\begin{figure}
%	\centering
%	\includegraphics[width=1\columnwidth]{cosinePDF.PNG}
%	\caption{Sample realization of the D2D network with directional antennas employed transmitters.}
%	\setlength{\abovecaptionskip}{-3cm}
%	\label{fig:cosinePDF}
%\end{figure}

\vspace{2mm}
\begin{lemma}\label{lem:lemma2}
	Given that the orientation AoA estimation error is modeled as a truncated Gaussian distribution as $\varepsilon \sim N_t (0,s^2, -\varepsilon_0, \varepsilon_0)$ and $| \varepsilon_0 |< \pi$, the probability distribution function (pdf) of D2D transmitter's antenna gain is
	\begin{align}\allowdisplaybreaks
	&f_{G}(g)=\frac{\sqrt{\zeta}\exp\left(-\zeta\arccos^2(\sqrt{\frac{g}{G_m}})\right)}{\text{erf}\left(\frac{\varepsilon_0}{\sqrt{2s^2}}\right) \sqrt{\pi}\sqrt{g}\sqrt{G_m-g}},\label{eq:pdfGauss}
	%F_G(g)&=
	\end{align}
	where
	\begin{align}\allowdisplaybreaks
	&\text{for} \hspace{2mm}  |\varepsilon_0|<\frac{\pi}{\tau}, \hspace{4mm}   g \in \left[\kappa G_m ,G_m\right]\nonumber\\
	&\text{for} \hspace{2mm} \frac{\pi}{\tau}\leq |\varepsilon_0|<\pi , \hspace{4mm}   g \in \left[0 ,G_m\right]\nonumber
	\end{align}
	and $\zeta=\frac{2}{\tau^2s^2}$.
\end{lemma}
\vspace{2mm}

\begin{IEEEproof}
	Following the same procedure as proof of Lemma \ref{lem:lemma1}, the antenna gain CDF with Gaussian distribution misalignment can be written as
	\begin{align}\allowdisplaybreaks
	%\text{for} \hspace{2mm}  |\varepsilon_0|<\frac{\pi}{\tau} \nonumber\\
	F_{G}(g)&=\bbP\left[G(\varepsilon)\leq g\right] \nonumber\\
	%&=\bbP\left[ \frac{2}{\tau}\arccos (\sqrt{\frac{g}{G_m}}) \leq |\varepsilon| \leq \varepsilon_0\right]\nonumber\\
	&=2\int_{G^{-1}(g)}^{\varepsilon_0} \frac{\exp(\frac{-\varepsilon^2}{2s^2})}{\sqrt{2\pi s^2}.\text{erf}\left(\frac{\varepsilon_0}{\sqrt{2s^2}}\right)}\text{d}\varepsilon\nonumber\\
	&= 1-\frac{\text{erf}\left(\sqrt{\zeta}\arccos(\sqrt{\frac{g}{G_m}})\right)}{\text{erf}\left(\frac{\varepsilon_0}{\sqrt{2s^2} }\right)}\nonumber
	\\
	F_G(g)&=
	\begin{cases}
	0 & g < 0\\
	1-\frac{\pi}{\tau \varepsilon_0} & g = 0 \\
	1-\frac{\text{erf}\left(\sqrt{\zeta}\arccos(\sqrt{\frac{g}{G_m}})\right)}{\text{erf}\left(\frac{\varepsilon_0}{\sqrt{2s^2} }\right)} & 0 < g \leq G_m\\
	1 & g > G_m \label{eq:CDFGuass}
	\end{cases}
	%\begin{cases}
	%0 & g < \kappa G_m \\
	%1-\frac{\text{erf}\left(\sqrt{\zeta}\arccos(\sqrt{\frac{g}{G_m}})\right)}{\text{erf}\left(\frac{\epsilon_0}{\sqrt{2} %s}\right)}\nonumber\\ & \kappa G_m \leq g \leq G_m\\
	%1 & g > G_m\\
	%\end{cases}
	%\\
	%\text{for} \hspace{2mm} \frac{\pi}{\tau}< |\varepsilon_0|<\pi &\nonumber\\
	%F_G(g)&=
	%\begin{cases}
	%0 & g < 0\\
	%2(1-\frac{\pi}{ \epsilon_0 \tau}) & g = 0\\
	%1-\frac{2}{\varepsilon_0\tau}\arccos \left(\sqrt{\frac{g}{G_m}}\right) & 0 < g \leq G_m\\
	%1 & g > G_m\label{eq:CDFGauss2}\\
	%\end{cases}
	\end{align}
\end{IEEEproof}
It is worth noting that beam alignment error would not change the distribution of interference, as the bore-sight angle of interferes are distributed uniformly and independently from the desired transmitter's signal. We will justify the accuracy of this assumption through simulations in Section \ref{sec:result}.

\section{Coverage Probability}\label{sec:coverageProb}

Using equations (\ref{eq:SINR}) and (\ref{eq:Inter}), and the antenna gain pdf in (\ref{eq:pdf}) and (\ref{eq:pdfGauss}), the SINR coverage probability for the typical receiver can be written as
%\begingroup\makeatletter\def\f@size{9}\check@mathfonts
\begin{align}\allowdisplaybreaks
p_\text{c}(\gamma) &= \bbP\left[\text{SINR}\geq \gamma\right]\nonumber\\
&=\bbP\left[\frac{h_0 G_0 (\varepsilon) d_0^{-\alpha}}{\sigma_n^2+I_n}\geq\gamma\right]\nonumber\\
&=\bbP \left[ h_0 \geq \frac{\gamma(\sigma_n^2+I_n)}{G_0 (\varepsilon) d_0^{-\alpha}}\right]\nonumber\\
&=\int \bbP \left[ h_0 \geq \frac{\gamma(\sigma_n^2+I_n)}{g_0 d_0^{-\alpha}}\big|G_0(\varepsilon)=g_0\right]f_{G_0}(g_0)\text{d}g_0 \nonumber\\
&= \int \bbE_{I_n}\left[e^{-\rho I_n}\right] e^{-\rho\sigma_n^2}f_{G_0}(g_0)\text{d}g_0\label{eq:SINRCovProb},
\end{align}
where $\sigma_n^2=\frac{\sigma^2}{P_D C}$ and $I_n=\frac{I}{P_D C}$ denote the normalized noise power and normalized aggregate interference, respectively. Notice that $\bbE[e^{-\rho I_n}]$ represents the Laplace transform of $I_n$ evaluated at $\rho=\frac{\gamma d_0^{\alpha}}{g_0}$ and can be written as
%\begingroup\makeatletter\def\f@size{9}\check@mathfonts
\begin{align}\allowdisplaybreaks
\mathcal{L}_{I_n}(\rho) &= \bbE_{I_n}\left[e^{-\rho I_n}\right]\nonumber\\
&=\bbE_{\mathbf{\Phi}_L}\left[e^{-\rho \sum_{i \in \mathbf{\Phi}_L} h_i G_i(\beta_i) \|x_i\|^{-\alpha}}\right]\nonumber\\
&=\bbE_{\mathbf{\Phi}_L}\left[\prod_{i \in \mathbf{\Phi}_L}\bbE_{h,G} \left[e^{-\rho h G \|x_i\|^{-\alpha}}\right]\right]\nonumber\\
&\overset{(a)}=e^{-2\pi \lambda_L \bbE_{h,G}\left[\int_{0}^{R}\left(1-e^{-\rho h G r^{-\alpha}}\right)r\text{d}r\right]}\nonumber\\
&=e^{-2\pi\lambda_L \chi(\rho)}\label{eq:Laplace},
\end{align}
where
%\begingroup\makeatletter\def\f@size{9}\check@mathfonts
\begin{align}\allowdisplaybreaks
\chi(\rho)&=
\frac{R^2}{2\tau}\left(1-{}_2F_1\left(-\delta,\frac{1}{2};1-\delta;-\rho R^{-\alpha}\right)\right)\nonumber\\
&\mathrel{\phantom{=}}-\frac{\Gamma(-\delta)\delta}{2\pi \rho^{-\delta}} \Gamma(1+\delta) \vartheta, \nonumber
\end{align}
where (a) follows due to moment generation function of Poisson distribution and also mapping the 2D Poisson point process, $\mathbf{\Phi}_L$ onto $\bbR^+$ by letting $\|x_i\|=r_i$ be the distances of points of $\mathbf{\Phi}_L$ from the typical receiver. $\delta=\frac{2}{\alpha_L}$, $\Gamma(z)=\int_{0}^{\infty}x^{z-1}e^{-x}\text{d}x$ represents the Gamma function, and ${}_2F_1(.)$ is the hyper-geometric function. Finally,  $\vartheta=\int_{0}^{\frac{\pi}{\tau}}G_m^{\delta}\cos^{2\delta}(\frac{\tau\theta}{2})$.
%\vspace{1mm}
\begin{IEEEproof}
Part of the Laplace transform of $I_n$, denoted by $\chi(\rho)$, can be written as
\begingroup\makeatletter\def\f@size{8}\check@mathfonts
\begin{align}\allowdisplaybreaks\label{eq:proof}
&\bbE_{h,G}\left[\int_{0}^{R}\left(1-e^{-\rho h G r^{-\alpha}}\right)r\text{d}r\right]\nonumber\\
&=\frac{R^2}{2}-\bbE_{h,G}\left[\frac{\delta}{2}(\rho hG)^{\delta}\Gamma(-\delta,\rho hGR^{-\alpha})\right]\nonumber\\
&\overset{(a)}=\frac{R^2}{2}-\frac{\delta }{2}\bbE_{h,G}\left[ \frac{\Gamma(-\delta)}{(\rho hG)^{-\delta}}- \sum_{k=0}^{\infty}\frac{(-\rho hGR^{-\alpha})^k R ^2} {k!(k-\delta)}\right]\nonumber  \\
&=\frac{R^2}{2}-\frac{\Gamma(-\delta)\delta}{2\rho ^{-\delta}}\bbE[h^{\delta}G^{\delta}]+\frac{\delta R^2}{2}\sum_{k=0}^{\infty}\frac{(-\rho R^{-\alpha})^k}{k!(k-\delta)}\bbE_{h,G}[h^kG^k]\nonumber\\
&\overset{(b)}=-\frac{\Gamma(-\delta)\delta}{2\pi \rho^{-\delta}} \Gamma(1+\delta)\int_{0}^{\frac{\pi}{\tau}}G_m^{\delta}\cos^{2\delta}(\frac{\tau \beta}{2})\text{d}\beta\nonumber\\
&\mathrel{\phantom{=}}+\frac{\delta R^2} {2\pi} \sum_{k=1}^{\infty}\frac{(-\rho R^{-\alpha})^k}{k!(k-\delta)}\Gamma(k+1) \int_{0}^{\frac{\pi}{\tau}}G_m^k\cos^{2k}( \frac{\tau \beta}{2})\text{d}\beta\nonumber\\
&=-\frac{\Gamma(-\delta)\delta}{2\pi \rho^{-\delta}} \Gamma(1+\delta) \vartheta\nonumber+\frac{\delta R^2}{2\tau} \sum_{k=1}^{\infty} \frac{(-\rho G_m R^{-\alpha}) ^k \Gamma(k+1)} {k!(k-\delta)} \frac{(2k)!}{4^k(k!)^2} \nonumber\\
&=-\frac{\Gamma(-\delta)\delta}{2\pi s^{-\delta}} \Gamma(1+\delta) \vartheta\nonumber+\frac{\delta R^2} {2\tau}\sum_{k=1}^{\infty}\frac{(-sR^{-\alpha})^k}{k!(k-\delta)}\frac{\Gamma(k+\frac{1}{2})}{\sqrt{\pi}}\nonumber\\
&=-\frac{\Gamma(-\delta)\delta}{2\pi \rho^{-\delta}} \Gamma(1+\delta) \vartheta\nonumber+\frac{R^2} {2\tau}\left(1- {}_2F_1\left(-\delta,\frac{1}{2};1-\delta;-\rho G_m R^{-\alpha}\right)\right)
\end{align}\endgroup
where $\Gamma(a,z)=\int_{z}^{\infty}t^{a-1}e^{-t}dt$ is the upper incomplete Gamma function. Step (a) is from the series expansion of $\Gamma(-\delta,shGR^{-\alpha})$. Step (b) follows due to the exponential distribution of channel gain, $h$, $\bbE[h^k]=\Gamma(k+1)$.
%where equation \eqref{eq:iid} follows since channel gains are i.i.d, and also PPP $\tilde{\mathbf{\Phi}}_\text{j}$ and $h_r$ are independent.
%Equations \eqref{eq:mgf_poisson} and \eqref{eq:mgf_exp} are derived using the probability generating functional (PGFL) of PPP $\tilde{\mathbf{\Phi}}_\text{j}$ with density $\lambda_{\mathbf{\Phi}_\text{j}}(r)$, and PGFL of $h_r$ with exponential distribution, respectively.
\end{IEEEproof}

\begin{table}
\caption{Simulation Parameters}
    \centering
       \begin{tabular}{|c|c|c|c}
       \hline
       Parameter&Notation& Value \\ \hline\hline\hline
       %BS/DT power & $P_B$, $P_{\text{D}}$ & $37$, $0$ (dBm) \\
       Antenna gain & $G_m$ & $10$ (dBi)  \\
       Mainlobe spread & $\tau$&  $1$,$2$,$3$,$4$\\
       Bore-sight angle & $\varphi_i$ & $\varphi \sim \mathcal{U}(-\pi,\pi)$\\
       %Error parameter & $\varepsilon_0$ & $0.3\pi$\\
       %Error distribution & $\varepsilon_U$ & $\varepsilon_U \sim \mathcal{U}(-\varepsilon_0,\varepsilon_0)$ \\
       %Error distribution & $\varepsilon_G$ & $\varepsilon_G \sim \mathcal{N}_t(0,s^2,-\varepsilon_0,\varepsilon_0)$ \\
       Density of PPP &  $\lambda$  & $50$ ($\text{km}^{-2}$)\\
       Radius of LOS ball & $R$ & $300$ (m)\\
       D2D TX power& $P_D$ & $1$ (watt)\\
       Path-loss exponent  & $\alpha$& $2.1$ \\
       Path-loss intercept  & $C$& $-62$ (dB) \\
       Bandwidth & $B$ & $1$ (GHz)\\
       Carrier frequency& $f$ & $28$ (GHz)\\
       %Simulation area & - & $10\times10$ ($km^2$)\\
       %Simulation iterations & - & $10,000$\\
       Noise power & $\sigma^2$ & $-174+10 \log_{10}B+10$ {\footnotesize(dBm)} \\

       \hline
       \end{tabular}\label{params}
\end{table}

%%%%%%%%%%%%%%%%%%%%%%%%%%%%%%%%%%%%%%%%%%%%%%%%%%%%%%%%%%%%%%%%%%%%%%%%%%%%%%%%%%%%%%%%%%%%%%%%%%
%%%%%%%%%%%%%%%%%%%%%%%%%%%%%%%%%%%%%%%%%%%% RESULTS %%%%%%%%%%%%%%%%%%%%%%%%%%%%%%%%%%%%%%%%%%%%%%
%%%%%%%%%%%%%%%%%%%%%%%%%%%%%%%%%%%%%%%%%%%%%%%%%%%%%%%%%%%%%%%%%%%%%%%%%%%%%%%%%%%%%%%%%%%%%%%%%%%

\section{Numerical Results and Discussions}\label{sec:result}
In this section, we evaluate the performance of the mmWave D2D network using the obtained expression for coverage probability in (\ref{eq:SINRCovProb}) and (\ref{eq:Laplace}). The impact of antenna misalignment on the network performance, due to inaccurate AoA estimation, is captured using the antenna gain distributions in the presence of the error in equations (\ref{eq:pdf}) and (\ref{eq:pdfGauss}).
Moreover, to validate our analytical results, we simulated a network similar to the one discussed in Section \ref{sec:system model}. For our simulations, we consider an area of the size $10$ $km$ $\times$ $10$ $km$  which is --given the transmit power of D2D devices-- large enough to avoid the boundary effect. D2D transmitters along with various size rectangular blockages are distributed in the area according to PPP. In addition, we assume that all the transmitters use a constant power for transmission. Table \ref{params} summarizes the simulation parameters. To thwart the effect of noisy data, we used Monte Carlo simulation with $10,000$ iterations and averaged out the results. In the following figures, simulation results are represented by "$+$" symbol.

 \begin{figure}
	\centering
	\includegraphics[width=.75\columnwidth, trim=4cm 8.5cm 4.0cm 9.1cm, clip]{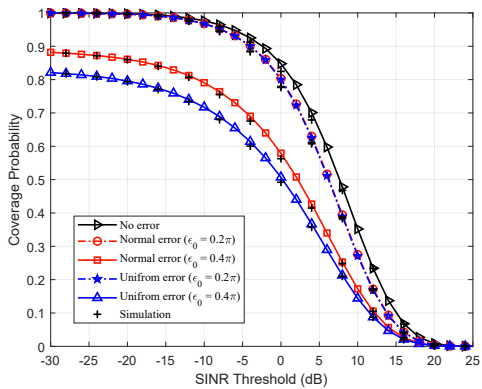}
	\caption{mmWave D2D network's SINR coverage probability vs. SINR threshold with $\tau = 3$ and $s^2=1$.}
	\label{fig:Coverage_error}
\end{figure}

% \begin{figure}
%\centering
%\includegraphics[width=.75\columnwidth, trim=4cm 8.5cm 4.0cm 9.1cm, clip]{coverageUGwidthC.pdf}
%\caption{mmWave D2D network's SINR coverage probability vs. SINR threshold with $s^2=1$.}
%	\label{fig:Coverage_width}
%\end{figure}

\begin{figure}
\centering
\includegraphics[width=.75\columnwidth, trim=4.1cm 8.5cm 4.0cm 9.1cm, clip]{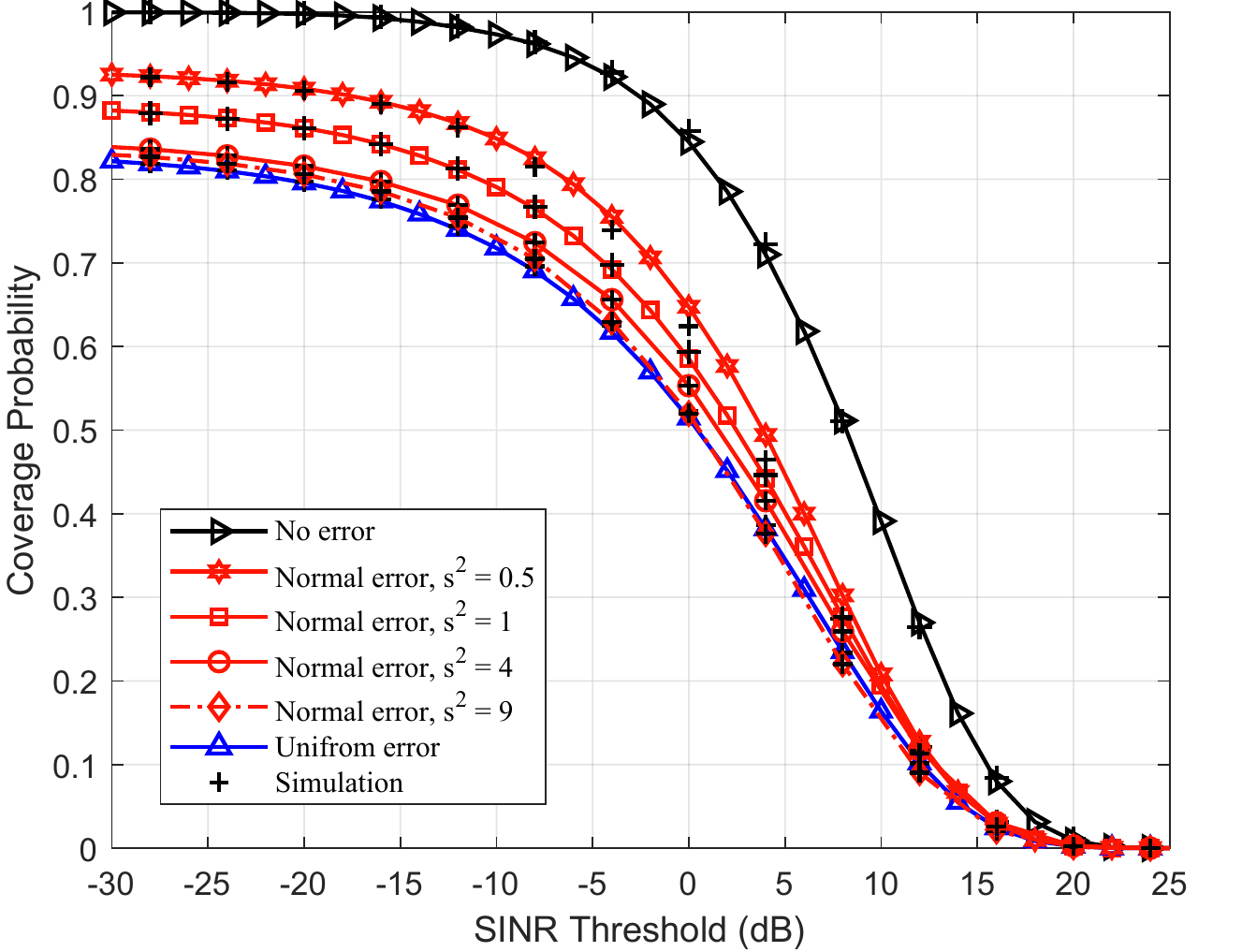}
\caption{mmWave D2D network's SINR coverage probability vs. SINR threshold with $\tau = 3$ and $\varepsilon_0 = 0.4 \pi$.}
	\label{fig:Cov-sigma}
\end{figure}

Figure \ref{fig:Coverage_error} %\ref{fig:Coverage_width}% 
and \ref{fig:Cov-sigma} show the SINR coverage probability of the directional D2D network in the mmWave band as a function of the SINR threshold, for three different scenarios, namely, perfect beam alignment, erroneous alignment with the Gaussian error distribution and erroneous alignment with uniform error distribution.

Figure \ref{fig:Coverage_error} demonstrates the impact of magnitude of error on the network coverage probability, with two different error magnitudes, i.e. $\varepsilon_0 = 0.2\pi$ and $\varepsilon_0 = 0.4 \pi$. It can be seen that the network coverage probability decreases as the error magnitude increases. Moreover, in case the ratio of error to antenna beam spread is bigger than one, $\frac{\varepsilon_0}{\pi/\tau}>1$, the chance of transmission with zero antenna gain increases which degrades the coverage probability significantly. This evaluation indicates that big alignment error has a significant impact on network performance, and should not be neglected as in \cite{yang2016analysis, thornburg2015ergodic}. Moreover, it is shown that the analytical results match the simulations with negligible gaps, which indicates the accuracy of equation (\ref{eq:SINRCovProb}).

%In Figure %\ref{fig:Coverage_width}%,
%we evaluated the impact of antenna beam spread on the network performance. In this simulation the ratio of the error magnitude to the antenna beam spread is fixed, i.e. $\frac{\varepsilon_0}{\pi/ \tau}=1.2$.  It is shown that in case of perfect beam alignment and erroneous alignment with uniform distribution, the narrower antenna provides better coverage as it reduces the amount of multi-user-interference (MUI). However, in the case of misalignment with a normal distribution, for small values of SINR threshold, the wider beam provides better coverage than the narrower one. The variance of the normal distribution is considered constant ($s^2=1$), hence the same amount of error leads to a greater reduction in the transmitter's antenna gain for the narrower beam. This, however, has almost no performance gain for higher SINR threshold.

In Figure \ref{fig:Cov-sigma}, the impact of variance of the normally distributed error is investigated on network's performance. As the variance of error increases the coverage probability decreases. This is mainly due to the higher chance of non-zero transmission gain at smaller variance. Larger values of variance leads to higher variation in transmission gain, which eventually cause the coverage probability degradation. Moreover, the graph shows that Gaussian distributed error with $s^2=9$ almost matches the uniformly distributed error. It is intuitive, as truncated Gaussian distribution with large variance resembles the uniform distribution.

%%%%%%%%%%%%%%%%%%%%%%%%%%%%%%%%%%%%%%%%%%%%%%%%%%%%%%%%%%%%%%%%%%%%%%%%%%%%%%%%%%%%%%%%%%%%%%%%%%%%%%%%%%%%%%%%%%%%%%%%%%%%%%%%%%%%
%%%%%%%%%%%%%%%%%%%%%%%%%%%%%%%%%%%%%%%%%%%%%%%%%%%%%%%%%%%%%%%%%%%%%%%%%%%%%%%%%%%%%%%%%%%%%%%%%%%%%%%%%%%%%%%%%%%%%%%%%%%%%%%%%%%%
%\setlength\belowcaptionskip{-6mm}
%\begin{figure}
%\centering
%\includegraphics[width=.81\columnwidth, trim=.8cm 1cm 1.5cm 1.5cm, clip]{rate-test.pdf}
%\caption{Rate coverage probability vs. achievable rate with $d_0$ $= 50$ $m$ and $\lambda_{\text{DT}}=50$ $km^{-2}$.}
%	\label{fig:rate}
%\end{figure}
%\setlength\belowcaptionskip{-7mm}

\begin{figure}
	\centering
	\includegraphics[width=.8\columnwidth, trim=3.9cm 8.5cm 4.0cm 9.1cm, clip]{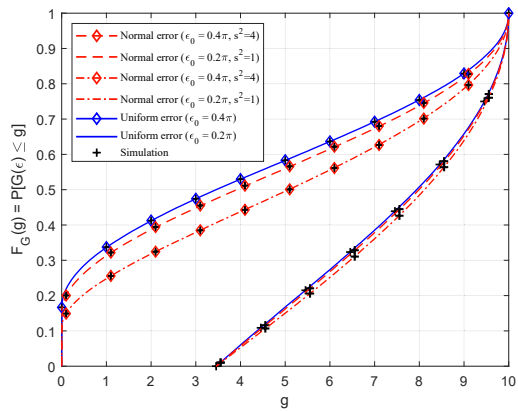}
	\caption{CDF of the antenna gain of the typical D2D transmitter in the directional mmWave D2D network with $\tau =3 $.}
	\label{fig:antGain}
\end{figure}

Figure \ref{fig:antGain} shows the CDF of D2D transmitters antenna gain with Gaussian and uniform error distribution, with different error magnitudes, i.e. $\varepsilon_0=0.2\pi$ and $\varepsilon_0=0.4\pi$. This simulation investigates the accuracy of Lemma \ref{lem:lemma1} and \ref{lem:lemma2}. It can be seen that for $\frac{\varepsilon_0}{\pi/\tau}<1$, antenna gain ranges from $\kappa G_m$ to $G_m$, while for $\frac{\varepsilon_0}{\pi/\tau}>1$ it changes from $0$ to $G_m$.

\begin{figure}
	\centering
	\includegraphics[width=.8\columnwidth, trim=3.9cm 8.5cm 4.2cm 9.1cm, clip]{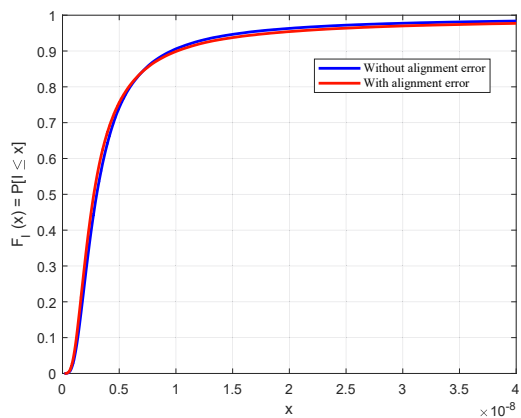}
	\caption{CDF of cumulative interference in a directional mmWave D2D network with $\tau=3$, $\varepsilon_0=0.4\pi$ and $\lambda=300$ users per $km^2$.}
	\label{fig:interf}
\end{figure}

Figure \ref{fig:interf} shows CDF of integrated interference in a directional D2D network with and without alignment error. As they are approximately matched, thus, our assumption on not considering the impact of the error on alignment is accurate.

\begin{figure}[h]
	\centering
	\includegraphics[width=.8\columnwidth, trim=4.1cm 8.5cm 4.2cm 9.1cm, clip]{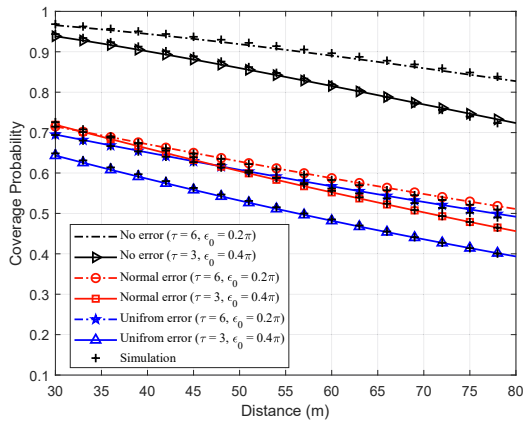}
	\caption{mmWave D2D network's SINR overage probability vs. D2D pair's distance with $\gamma = -5 $ (dB).}
	\label{fig:distance}
\end{figure}
Figure \ref{fig:distance} shows SINR coverage probability of D2D network as a function of the distance among D2D transmitter-receiver pairs for SINR threshold $\gamma = -5$ dB. It is shown that, increasing
the distance of D2D pairs, degrades the performance of the D2D network with and without misalignment. However, in the presence of the misalignment increasing the distance drops the network performance even more, due to the decrease in transmitter antenna gain.

\section{Conclusion and Future Work}\label{sec:Conclusion}
In this paper, we proposed a mathematical framework to analyze the impact of AoA estimation on the performance of a mmWave D2D network. Based on the prior information we have about the error, the AoA estimation error is modeled using normal and uniform distributions.
We have used stochastic geometry to provide a complete framework to analyze the D2D network performance in the presence of error in terms of the received SINR coverage probability, for which analytical formulas are derived. Simulation results show that the coverage of the network with erroneous beam alignment can be degraded by about $35\%$ compared to the one with perfect beam alignment. Moreover, our simulations validate the analytical results discussed in the paper. Considering the significant impact of beam alignment error on the network performance, proposing a mechanism that corrects and compensate the beam alignment error using a feedback loop is a promising future direction.

\medskip

\bibliographystyle{ieeetr}
\bibliography{bib}

\end{document}